\documentclass[twocolumn,apl]{revtex4-1}
\usepackage{amssymb,amsmath}
\usepackage{graphicx}
\usepackage{dcolumn}
\usepackage{multirow}
\usepackage{color}
\definecolor{purple}{rgb}{0.61,0.19,1.00}
\usepackage{hyperref}

\newcommand{\drw}[1]{\textcolor{black}{#1}}
\newcommand{\snc}[1]{\textcolor{black}{#1}}
\newcommand{\DRW}[1]{\textcolor{black}{#1}}

\begin{document}

\title{\textbf{\fontfamily{phv}\selectfont Integration of on-chip \DRW{field-effect transistor} switches with dopantless Si/SiGe quantum dots for high-throughput testing}}
\author{Daniel R. Ward$^1$}
\author{D. E. Savage$^2$}
\author{M. G. Lagally$^2$}
\author{S. N. Coppersmith$^1$}
\author{M. A. Eriksson$^1$}
\affiliation{$^1$Department of Physics, University of Wisconsin-Madison, Madison, WI 53706, \DRW{USA}
\\$^2$Department of Materials Science \& Engineering, University of Wisconsin-Madison, Madison, WI 53706, \DRW{USA}}

\begin{abstract}
Measurement of multiple quantum devices on a single chip increases characterization throughput and enables testing of device repeatability, process yield, and systematic variations in device design.  We present a method that \snc{uses} on-chip field-effect transistor switches to enable multiplexed \snc{cryogenic} measurements of double quantum dot Si/SiGe devices.  Multiplexing
\snc{makes it feasible to characterize a number of devices that scales exponentially
with the number of external wires, a key capability given the significant constraints on
cryostat wiring currently in common use}.  We use this approach to characterize three nominally identical \DRW{quantum-point contact}
channels, enabling comparison of their threshold voltages for accumulation and their pinch-off voltages during a single cool-down of a dilution refrigerator.  
\end{abstract}

\maketitle


The fabrication of quantum dots in semiconductor heterostructures is challenging.  While it is common to make many devices in parallel, typically only one is wired up for a given dilution refrigerator cool-down.  Because there is statistical variation in the quality of the devices, there is no reason to believe that the first device which shows some level of proper functioning is the best device available.  Nonetheless, because \snc{taking a working sample out of a cryostat involves time and risk}, in practice the first working device is often the only one that is measured.  While \snc{much} progress has been made using the conventional approach of measuring one device per chip
\cite{Petta:2005p2180,Koppens:2006p766,Gustavsson:NanoLett2008,Simmons:2011p156804,Maune:2012p344},  
\snc{testing multiple devices during a single crystat cool-down
would significantly enhance efficiency.}  Furthermore, measuring multiple samples during a single cool-down would enable studying systematic variations in device design.

Semiconductor quantum devices are becoming increasingly complex, driven in part by a desire to develop and understand the large variety of qubits such structures can host~\cite{Loss:1998p120,Kane:1998p133,Levy:2002p1446,Hollenberg:2006p1534,Shi:2012p140503}.  
While simple device designs in doped heterostructures require only a few gates~\cite{Simmons:2007p213103}, 
 the recent \snc{transitions} to accumulation-mode, undoped devices \cite{borselliAPL2011,Maune:2012p344} 
\snc{and to}
more complicated depletion gate designs~\cite{Prance:2012p046808} 
\snc{necessitate many more} electrical connections per device.  \snc{The 
 devices discussed} below \snc{each have 22 connections (11 depletion gates, 5 accumulation gates, and 6 ohmic contacts)}; simply wiring multiple devices in parallel would require more wires than are conventionally available in low-temperature cryostats.  
In this context, on-chip multiplexing \snc{enables} convenient, low-risk testing of multiple devices per chip and per cryostat cool-down.

\snc{This letter presents} a method to improve the throughput for testing quantum devices.  We use undoped, accumulation-mode Si/SiGe with two layers of electrostatic gates~\cite{borselliAPL2011}.  We present a method for wiring, cooling, and measurement of four double dot structures at a time using integrated \DRW{field-effect transistor (FET)}
switches on the sample.
\snc{Our heterostructures are undoped and have a positive threshold for accumulation, so each quantum device contains no free carriers unless a positive voltage is applied to the accumulation gates.
Therefore,}
multiplexing solely the accumulation gates 
\snc{enables} 
independent measurement of all
\snc{the} quantum devices.  The multiplexer makes use of $2n$ \drw{switch control lines} for $2^{n} $ dot structures \cite{HorowitzAndHill}.
The multiplexed, four-device chip we describe here is controlled with a total of  22 DC lines for ohmic contacts, depletion gates, and accumulation gates, and an additional 4 DC switch \drw{control} lines are used for the multiplexer, resulting in a total of 26 connections.  This can be compared to the 37 connections that would be required without the multiplexer.  The scaling benefits of the multiplexing approach 
\snc{increase exponentially with $n$}.


The method \snc{presented here} should be \snc{applicable to accumulation-mode} quantum devices made from any semiconductor-based two-dimensional electron system, including Si metal-oxide-semiconductor approaches \cite{Angus:2008p112103,Xiao:2010p1876}, donor-based devices \cite{morelloNature10}, and non-modulation doped GaAs \cite{See2010APL}.  Our method may be extended to other semiconductor heterostructures where depletion or accumulation mode switches are possible --- such as doped Si/SiGe, GaAs, or InAs \cite{simmonsNL09,Elzerman:2003p728,schroerNanoLett2010} --- 
by adding
a second layer of gates to those devices.  

%
%
\begin{figure*}[btph]
\includegraphics[width=1.9\columnwidth]{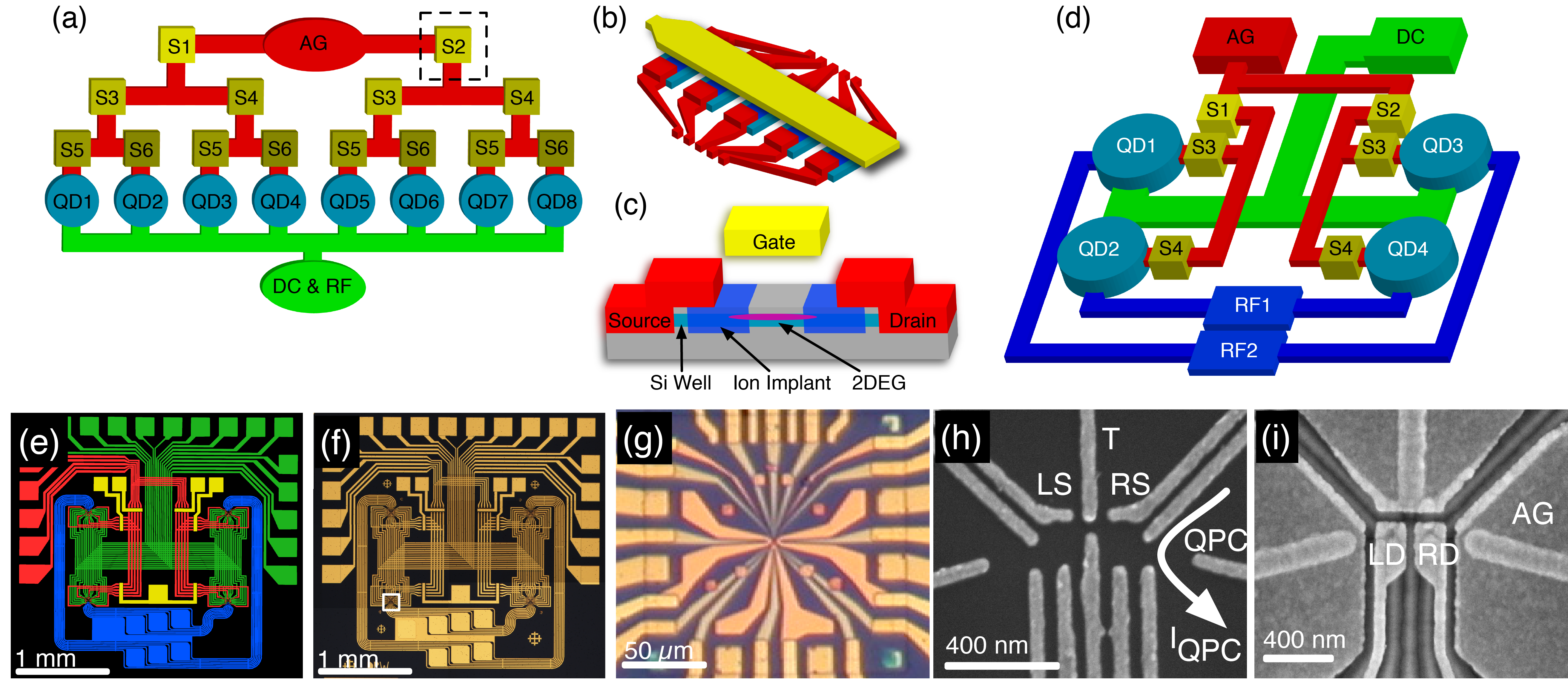}
\caption{\label{fig1} 
Schematic of multiplexing theory and implementation.  
(a) Schematic of multiplexing \snc{of eight accumulation-mode  
undoped Si/SiGe double quantum dots.  The hierarchical structure can be extended straightforwardly
to multiplex $2^n$ devices with a number of external lines that grows linearly with $n$.}  
All the DC and RF lines (green) are connected in parallel to every QD (blue).  AG represents an \snc{electrical} bus for five accumulation gate control lines (red) which pass through a series of branching switches \drw{blocks} (yellows).  Control lines are shared by switches with the same label.  
(b) Schematic of the outlined switch block S2 from (a).  The accumulation gate bus
(red) is routed through five on-chip FET heterostructures (blue) in parallel that share a common gate electrode (yellow).
(c)  Cross-section of an on-chip FET switch seen in (b).  Source and drain leads come onto the raised mesa of active heterostructure.  Electrical connections to the strained silicon well are made through ion-implanted regions (dark blue) on the mesa.  
\snc{Conduction between the source and drain occurs only when the top gate voltage exceeds a positive threshold at which a 2DEG forms in the Si well.}
(d) Schematic of the 
physical layout for four multiplexed, double quantum dot structures using the same color scheme as in (a).  The RF lines (dark blue) are kept in a separate bus that 
\snc{never overlaps} any of the other buses or itself\snc{,} to minimize cross-talk.  
(e) False color image of actual multiplexed, undoped Si/SiGe accumulation mode device using the same color scheme as (a).
(f) Same as in (e) without the color modification.  The small white box in the lower left indicates 
where the bottom-left device is located.
(g) Optical image of the bottom left device from the boxed region in (f).
(h)  SEM image of the depletion gates of the quantum device (image acquired partway through the fabrication process).
(i)  SEM image of a completed quantum device\snc{,} showing the accumulation gates on top of 80~nm of Al$_{2}$O$_{3}$; the depletion gates from (h) can be observed in the background of the image.  }
\end{figure*}

Figure~1(a) illustrates how eight double quantum dots could be multiplexed using 14 switch blocks (yellow) and six switch control lines (labeled S1-S6) wired to multiple switches in parallel.  The accumulation gate bus (red) consists of 5 accumulation gate voltage lines that run between the switch blocks and finally to each double dot structure.  Accumulation gate voltages can be applied to a single double dot structure by applying positive voltage to the correct switch control lines.  For instance, 
accumulation gate voltages \snc{can be applied} only to device QD1 \snc{by switching control lines S1, S3, and S5 to positive voltage} while \snc{ leaving S2, S4, and S6 at zero or negative voltage}. 
\snc{Current flows through a structure only if its accumulation gate voltage is positive,}
so all  
DC and RF gates (green)
\snc{and} ohmic connections 
\snc{can} be wired in parallel to every double dot structure.  
\DRW{We do not expect the presence of the FET network to have significant influence on the behavior of the quantum devices being multiplexed.  The primary side effect of the network will be to increase by a  few k$\Omega$ the resistance of the electrical path leading to each of the multiplexed accumulation gates, which is not expected to affect the gate performance. 
}

A switch block consists of 5 individual switches, one for each accumulation gate line, controlled by \snc{a} shared gate, as shown in Fig.~1(b).  The switches in the switch block work like traditional accumulation-mode FET switches, 
\snc{except that}
electrons are confined to \snc{a two-dimensional electron gas (2DEG)}.  Fig.~1(c) illustrates the basic layout of single switch.  A bias is applied to the left ohmic contact, which \snc{in the multiplexed four-device structure that we fabricated} is embedded in the 20$\times$80~$\mu$m$^2$ mesa of active heterostructure.  An accumulation gate, overlapping the right and left ohmic contacts (18$\times$20~$\mu$m$^2$) as well as the material in between, is biased positive relative to the left ohmic contact to induce a 2DEG between the contacts\snc{.
When 
the switch is off,} 
the isolation is better than 10~G$\Omega$.  
\snc{Because} the gate voltage must be referenced relative to the ohmic contacts, the switches must operate at approximately twice the accumulation threshold of the material, since the accumulation gates themselves operate above the threshold voltage.  

\snc{Fig.~1(d) shows the
layout of 
a four double-dot device}.  The three electrical buses provide access to (i) DC gate and ohmic contacts, (ii) accumulation gates, and (iii) RF connections to 
\snc{some} of the gates.  The accumulation-gate bus is multiplexed by the six switch blocks connected to the four switch control lines S1-S4.  To minimize RF cross-talk, the RF gates 
\snc{are} separated from the other gates and \snc{do not} overlap any other structures\snc{, which requires}
one RF bus per two devices.

\snc{Fig.}~1(e,f) show a false color image and the unmodified image, respectively, of a \snc{fabricated} device with four multiplexed double quantum dots, each with two integrated charge sensors.  The colors in Fig.~1(e) correspond to the various buses from Fig.~1(d).
\snc{The device is fabricated in a Si/SiGe heterostructure} grown \snc{using} chemical vapor deposition 
on a SiGe (001) substrate.  A 800~nm Si$_{0.7}$Ge$_{0.3}$ buffer is deposited\snc{,} followed by a 12~nm thick strained Si well\snc{.} 
\snc{A} 32~nm Si$_{0.7}$Ge$_{0.3}$ layer is \snc{then} deposited\snc{,} followed by a 1~nm thick Si cap layer.  To avoid unwanted accumulation and leakage\snc{, most} of the substrate is etched below the Si well with reactive ion etching, leaving active material in small 100$\times$100~$\mu$m$^2$ mesas for the dot structures \snc{and} 
20$\times$80~$\mu$m$^2$ mesas 
\snc{for the 
switches}.  All exposed surfaces are then coated in 10~nm of Al$_2$O$_3$ via atomic layer deposition (ALD).  Ohmic contacts to the 2DEG in the dot structures and switches are created by 20~kV phosphorous implantation activated with a 15~s, 700$^{\circ}$C anneal.  
\snc{Two layers of gates, separated by an isolating layer of 80~nm of Al$_2$0$_3$ deposited by ALD, are defined by a combination of photo- and electron-beam lithography and deposited by electron-beam evaporation of 1.7~nm~Ti/40~nm~Au.}    
\snc{Interconnects between the two layers of gates are made} by etching vias in the Al$_2$O$_3$ with dilute HF before depositing the second layer of gates.  \snc{Fig.~1(g) shows a single double dot structure, and Figs.~1(h,i) show the fine features of the double dot gate structure}.

%
%
\begin{figure}[]
\includegraphics[width=0.45\textwidth]{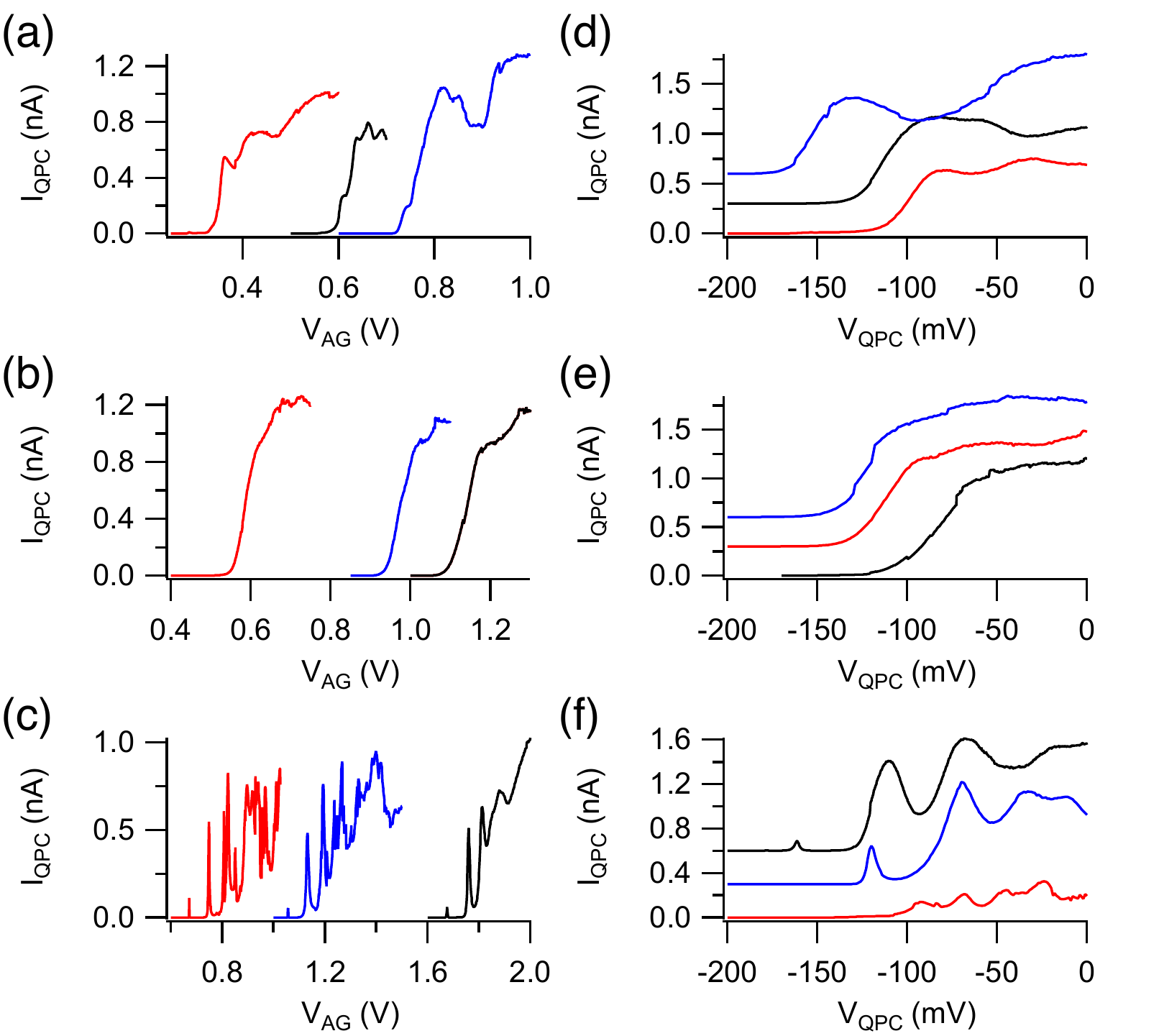}
\caption{\label{fig2} 
(a--c) QPC current $I_{QPC}$ ($V_{SD} = 43 \mu V$) as a function of the QPC accumulation gate voltage $V_{AG}$ for the top-left (a), bottom-left (b), and bottom-right (c) dot structures of Fig. 1a.    Each curve represents a different trial (in the order black, blue, red) after illuminating the sample.  
\snc{While the threshold} voltage for accumulation shifts for each trial, the turn\snc{-}on behavior (the number and sharpness of the peaks in the curves) is qualitatively similar for and characteristic of each specific QPC channel, analogous to a device fingerprint.
(d--f) QPC current $I_{QPC}$ ($V_{SD} = 43 \mu V$) as a function of the QPC depletion gate voltage $V_{QPC}$ for the top-left (a), bottom-left (b), and bottom-right (c) devices, 
\snc{ taken
at the accumulation gate voltages as described in the text}.  Curves have been offset by 0.3~nA for clarity.  Each curve represents a different trial (in the order black, blue, red) after illuminating the sample.  We again observe that the shape of the curves is similar in each trial for a given device.
}
\end{figure}

To test our multiplexed sample design\snc{, we measured} the accumulation threshold and pinch\snc{-}off characteristics of a \DRW{quantum point contact (QPC)}
channel (labelled QPC in Fig.~1(h)) on three of the four dot structures \snc{ during  one} refrigerator cooldown \textcolor{black}{to $\sim200$~mK.}  \DRW{The fourth dot structure could not be measured due to a failed wire bond for one of the multiplexing switches.}   To measure the accumulation threshold, \snc{we applied a small \snc{~43}~$\mu$V bias across the ohmics and measured current flow through the channel while increasing the accumulation gate voltage (AG in Fig.~1(i)) over the QPC channel, with all other gates held at zero}.   For each device we repeated the measurement three times, illuminating \textcolor{black}{ the sample for 20~s with a laser diode} between tests to reset the device.  As shown in Fig.~2(a--c), while the accumulation threshold for a device shifts after resetting, the character of the curve 
\snc{(the relative location of peaks and the number of peaks and dips)}
remains similar.  
\snc{The differences in the accumulation behaviors 
indicate that} nanoscale differences in gate geometry or heterostructure and material defects 
\snc{significantly affect} device performance and that these effects are robust against repeated illumination.

We also tested QPC pinch\snc{-}off in all three channels, again repeating the experiment with a reset of the device between measurements.   Here\snc{,} 
we start with 
the QPC gate 
at zero volts and the accumulation gate at \snc{a} point on the accumulation curve just past where accumulation levels 
\drw{become stable, ranging from 0.31~V above accumulation threshold for the top-left device to 0.9~V for for the bottom-right device.}
We then \snc{sweep} the QPC gate negative to pinch off the conduction channel.  As shown in Fig.~2(d--f), each of the three channels is unique.  After testing, the structures were imaged in a scanning electron microscope.
\snc{No} discernible differences were visible between the devices, suggesting that the observed differences in device properties arise from microscopic differences in the heterostructure or gate dielectrics.

We have presented a working implementation of a multiplexed structure for high\snc{-}throughput testing of multiple quantum dot structures on a single chip using integrated FET switches.  
\snc{Multiple} devices can be quickly screened for the desired characteristics 
and then the best device can be further tested without the time-consuming and potentially sample-ruining process of inserting and removing samples from the dilution refrigerator.  Further, we have demonstrated integration of classical electrical components on-chip with quantum devices.  

\DRW{We thank R. Mohr for discussions regarding ion-implantation.}   \snc{This work was supported in part by the U.S. Army Research Office (W911NF-08-1-0482, W911NF-12-1-0607) and by the United States Department of Defense.  The views and conclusions
contained in this document are those of the authors and should not be interpreted as representing the
official policies, either expressly or implied, of the US Government.  Development and maintenance of the growth facilities used for fabricating samples is supported by DOE (DE-FG02-03ER46028).}


%

\end{document}